\documentclass[12pt]{iopart}

\usepackage[utf8]{inputenc}  

	\usepackage[T1]{fontenc}  

\expandafter\let\csname equation*\endcsname\relax
  \expandafter\let\csname endequation*\endcsname\relax

	\usepackage[intlimits]{amsmath}		

	\usepackage{amssymb}

	\usepackage{bbold}

	\usepackage{dsfont}      

	\usepackage{fixmath}     

\usepackage{caption}
	\usepackage{graphicx}

	\usepackage{ifthen}
	
	\usepackage{epstopdf}

	\usepackage{braket}   
	
	\usepackage[babel,english=american]{csquotes} 

	\usepackage[final,colorlinks=true,linkcolor=blue,citecolor=blue]{hyperref}    

  \newcommand{\ii}{\mathrm{i}}

  \newcommand{\abs}[1]{\left\lvert{#1}\right\rvert}
  
  \newcommand*{\defeq}{\mathrel{\vcenter{\baselineskip0.5ex \lineskiplimit0pt
                     \hbox{\scriptsize.}\hbox{\scriptsize.}}}%
                     =}

  \newlength{\bracewidth}

	\newcommand*\Einchar{E}
	\newcommand*\Einplus[2]{\hat{\Einchar}^{(+)}_{#1}\ifthenelse{\equal{\unexpanded{#2}}{}}{}{(\tout{#2})}}
	\newcommand*\Einminus[2]{\hat{\Einchar}^{(-)}_{#1}\ifthenelse{\equal{\unexpanded{#2}}{}}{}{(\tout{#2})}}

\newcommand\tout[1]{t_{#1}}

\newcommand{\aref}[1]{\hyperref[1]{Appendix~\ref{#1}}}
\newcommand{\difd}{\mathrm{d}}

\newcommand{\avg}[1]{\left\langle #1 \right\rangle}

\renewcommand{\abs}[1]{\left| #1 \right|}
\newcommand{\abssq}[1]{\left| #1 \right|^2}

\makeatletter
\newcommand*{\da@rightarrow}{\mathchar"0\hexnumber@\symAMSa 4B }
\newcommand*{\da@leftarrow}{\mathchar"0\hexnumber@\symAMSa 4C }
\newcommand*{\xdashrightarrow}[2][]{%
  \mathrel{%
    \mathpalette{\da@xarrow{#1}{#2}{}\da@rightarrow{\,}{}}{}%
  }%
}
\newcommand{\xdashleftarrow}[2][]{%
  \mathrel{%
    \mathpalette{\da@xarrow{#1}{#2}\da@leftarrow{}{}{\,}}{}%
  }%
}
\newcommand*{\da@xarrow}[7]{%
  \sbox0{$\ifx#7\scriptstyle\scriptscriptstyle\else\scriptstyle\fi#5#1#6\m@th$}%
  \sbox2{$\ifx#7\scriptstyle\scriptscriptstyle\else\scriptstyle\fi#5#2#6\m@th$}%
  \sbox4{$#7\dabar@\m@th$}%
  \dimen@=\wd0 %
  \ifdim\wd2 >\dimen@
    \dimen@=\wd2 %
  \fi
  \count@=2 %
  \def\da@bars{\dabar@\dabar@}%
  \@whiledim\count@\wd4<\dimen@\do{%
    \advance\count@\@ne
    \expandafter\def\expandafter\da@bars\expandafter{%
      \da@bars
      \dabar@ 
    }%
  }%
  \mathrel{#3}%
  \mathrel{%
    \mathop{\da@bars}\limits
    \ifx\\#1\\%
    \else
      _{\copy0}%
    \fi
    \ifx\\#2\\%
    \else
      ^{\copy2}%
    \fi
  }%
  \mathrel{#4}%
}
\makeatother

\usepackage{xypic,color}
\usepackage{tikz}
\newcommand{\xdotarrow}[2][->]{
\tikz[baseline=-\the\dimexpr\fontdimen22\textfont2\relax]{
\node[anchor=south,font=\scriptsize, inner ysep=.5pt,outer xsep=2.2pt](x){#2};
\color{blue};
\draw[shorten <=1.4pt,shorten >=1.4pt,dotted,#1](x.south west)--(x.south east);
}
}

\usepackage{tikz}
\newcommand{\xdasharrow}[2][->]{
\tikz[baseline=-\the\dimexpr\fontdimen22\textfont2\relax]{
\node[anchor=south,font=\scriptsize, inner ysep=.5pt,outer xsep=2.2pt](x){#2};
\color{red};
\draw[shorten <=1.4pt,shorten >=1.4pt,dashed,#1](x.south west)--(x.south east);
}
}
\begin{document}

\title{The physics of thermal light second-order interference beyond coherence}

\author{Vincenzo Tamma}

\address{School of Mathematics and Physics, University of Portsmouth, Portsmouth PO1 3QL, UK}
\address{Institute of Cosmology \& Gravitation, University of Portsmouth, Portsmouth PO1 3FX, UK}
\ead{vincenzo.tamma@port.ac.uk}
\vspace{10pt}
\begin{indented}
\item[]June 2018
\end{indented}

\begin{abstract}
A novel thermal light interferometer was recently introduced in Ref. \cite{Tam-Sei}. Here, two classically correlated beams, obtained by beam splitting a thermal light beam, propagate through two unbalanced Mach-Zehnder interferometers. Remarkably, second-order interference  between the long and the short paths in the two interferometers was predicted independently of how far, in principle, the length difference between the long and short paths is beyond the coherence length of the source. This phenomenon seems to contradict our common understanding of second-order coherence. We provide here a simple description of the physics underlying this effect in terms of two-photon interference.
\end{abstract}

%
%
%
%
%

\section{Introduction and motivation}
The	discover of the Hanbury Brown and Twiss (HBT) effect \cite{HBT1, HBT2}  in $1956$ triggered an intense debate about the physics behind this effect and even its correctness. To quote Hanbury Brown, the reason behind a lot of this criticism was that ``many physicists had failed to grasp how profoundly mysterious light really is''. However, as often happens in science for a new phenomenon which is initially not well understood and received, the HBT effect paved the way to the development of an entire new field, the field of quantum optics \cite{Schleich,Glauber2007,Scully1997,ShihBook,Mandel1995}. One can consider this discovery as a precursor of fundamental studies and experiments based on multi-photon interference and correlations \cite{Glauber1963,GlauberLecture,ShihAlley, HOM, Franson1989, Legero,HOMspatial, LiuSpatial, tammalaibacher2014a, tammalaibacher2014,Tam-Sei,Kim}, with intriguing applications in  imaging  \cite{Genov,  Lemos, Jane, PittShi, ValScaD'AnShi, D'AnShi, Scarcelli,  gatti2, ferri, Chen, Luo}, quantum information processing \cite{Kok, nielsen,BosonSampling, laibachertamma2015, tammalaibacher2015b,tammalaibacher2015}, and metrology \cite{GioLloMac1, GioLloMac, Dowling, Cassano, DAngelo, Dowling2, Zimmermann, NOON}.

In the temporal domain \cite{HBT1}, the HBT effect consists of beam splitting a thermal beam and  measuring the correlation in time between the photon intensities at the beam splitter output ports (Fig. 1(a)). Two-photon interference is observed  for a detection time delay small compared to the coherence time of the source. Indeed,  two-fold detection events have twice the chance to occur at equal detection times than with a time delay much larger than the coherence time.  This effect arises from the interference between the two possible indistinguishable  pairs of paths that two detected photons  may have undertaken from the source to the two detectors (Fig. 1(b)).  We emphasize that, since these two pairs of path completely overlap in space, the HBT  effect is not sensitive to any relative phase delay. Indeed, no sinusoidal fringes can be observed in standard HBT experiments.

\begin{figure}
\centering
\captionsetup{justification=centering}
\includegraphics[scale=.5]{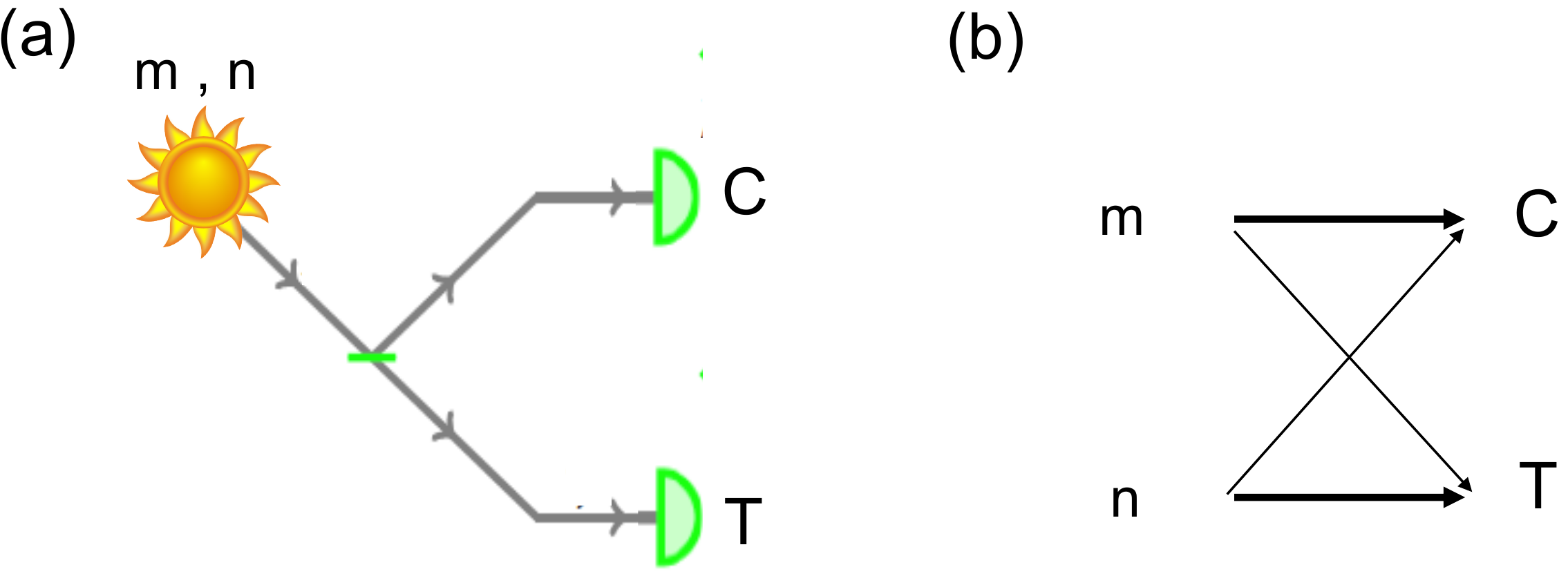}
\captionsetup{justification=justified}
\caption{(a) Hanbury Brown and Twiss (HBT) interferometer: The classically correlated beams obtained by beam splitting the light emitted by a thermal source are measured by performing joint detections at the two output ports of the balanced beam splitter. (b) Two-photon interference is observed at detection time delays small compared with the coherence time of the source. This corresponds to the interference of two two-photon detection amplitudes corresponding to the two possible indistinguishable paths (depicted with bold and thin arrows, respectively) two photons m and n can undertake to trigger a joint detection. }

\label{fig:3}
\end{figure}

\begin{figure}
\centering
\captionsetup{justification=centering}
\includegraphics[scale=.5]{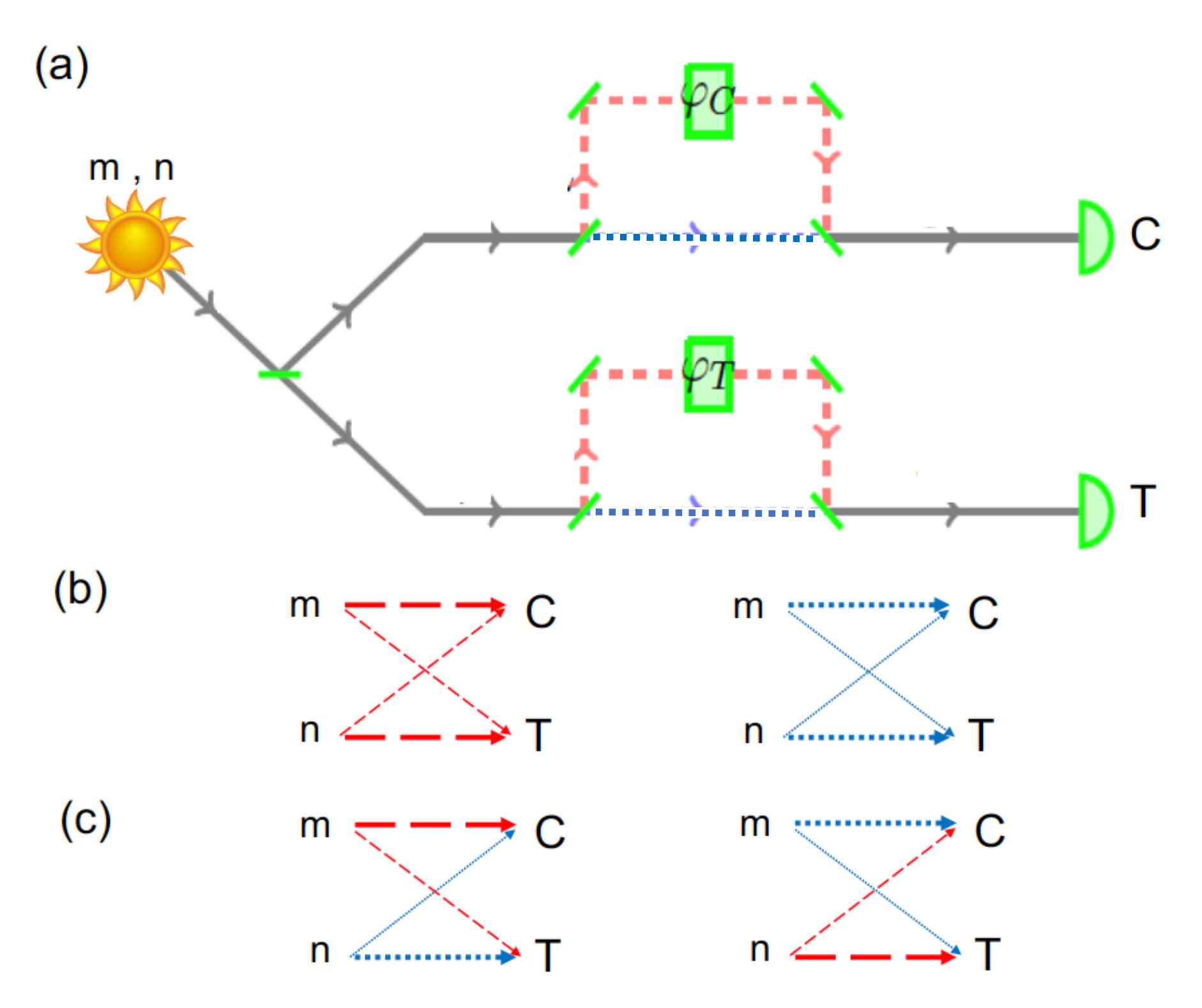}
\captionsetup{justification=justified}
\caption{ (a) Thermal light interferometer introduced in Ref. \cite{Tam-Sei}: A source of classically correlated beams as in the HBT interferometer in Fig. 1(a) is used. However, correlated measurements are not performed directly at the two output ports of the beam splitter but after the light propagates through two unbalanced Mach-Zehnder interferometers with relative phases $\varphi_C$ and $\varphi_T$, respectively. At detection time delays small compared with the coherence time of the source, second-order interference is observed  between the pair of long (dashed red) paths $L=L_C,L_T$ and the pair of short (dotted blue) paths $S=S_C,S_T$, which length difference $L-S$ is beyond the coherence length of the source. This interference is a result of the collective contribution of all the possible ways two-photon interference can arise between two pairs of two-photon paths (indicated by bold and thin arrows, respectively) undertaken by any given pair $(m,n)$ of detected photons emitted  by the thermal source: (b) two-photon interference between the $(L_C,L_T)$ and $(L_T,L_C)$ (or  $(S_C,S_T)$ and $(S_T,S_C)$) pairs of paths, which is phase-independent   as in standard HBT experiments;  (c) two-photon interference between the $(L_C,S_T)$ and $(L_T,S_C)$ (or  $(S_C,L_T)$ and $(S_T,L_C)$) pairs of paths, which  is instead sensitive to the phase difference $\varphi_C-\varphi_T$.}
\label{fig:3}
\end{figure}

Recently, was proposed a novel interferometric scheme, depicted in Fig. 2(a), where, differently from the standard HBT scheme, two Mach-Zehnder interferometers are introduced at the two output ports of the beam splitter before a correlation measurement at approximately equal detection times is performed \cite{Tam-Sei}.
Furthermore, the difference between the long (dashed red) paths $L=L_C,L_T$ and the short (dotted blue) paths $S=S_C,S_T$  in each Mach-Zehnder interferometer is assumed to be larger than the coherence length of the source, while the differences in length between paths of the same type, $L_C-L_T$ and $S_C-S_T$, are negligible with respect to the coherence length. 

 This interferometer has some similarities with the famous Franson interferometer \cite{Franson1989}, which however differs in two fundamental aspects:

-	a source of two classically correlated beams obtained by beam splitting a thermal light beam is replaced in the Franson interferometer  by a source of two photons entangled in energy and time;

-	the time delay between the long and short paths in each Mach-Zehnder interferometer is required to be within the second-order coherence time of the two-photon source, which is usually given by the coherence time of the pump laser which generates the entangled pair of photons when interacting with an SPDC crystal.

These two conditions ensure the emission of two entangled photons simultaneously but with a ``quantum uncertainty'' in time (second-order coherence time) larger than the time delay $(L-S)/c$ due to the unbalance between the long and short paths in each Mach-Zehnder interferometer, even if the usual first order coherence length of each single photon is much smaller than the path length difference  $L-S$.  This is at the very heart of the observation of the well known two-photon Franson interference between the two pairs $(L_C,L_T)$ and $(S_C,S_T)$ of two-photon paths. 


Can second order interference between the path configurations $(L_C,L_T)$ and $(S_C,S_T)$  be also observed in the  thermal light interferometer in Fig. 2(a)  where  the path length difference $\vert L -S\vert/c$  in each Mach-Zehnder interferometer is beyond the coherence length of the thermal source? Does it matter here how far beyond?  Interestingly, as shown in Ref. \cite{Tam-Sei}, the answer is \textit{yes} and it does \textit{not} matter, in principle, how far beyond. This result is highly counter-intuitive since, differently from the Franson interferometer, no energy-time entanglement  occurs to justify any two-path indistinguishability  in terms of second-order coherence length. How does such two-path indistinguishability  arise?    How can we reconcile it with our common understanding of second-order coherence? 

Even more interestingly, it was predicted that this second-order interference phenomenon with thermal light, differently from the standard HBT effect, could manifest itself in sinusoidal fringes as a function of the \textit{difference} $\varphi_{C}-\varphi_{T}$ between the phase delays in the two Mach-Zehnder interferometer  \cite{Tam-Sei}. This differs from the Franson interferometer which is sensitive to the \textit{sum} $\varphi_{C}+\varphi_{T}$ of the interferometer relative phases. 

But what is the physics behind the observation of these sinusoidal fringes? And why such a fundamental difference with respect to the Franson interferometer?

All these questions  have puzzled different scientists who were reluctant in believing that this interference effect was even correct when it was first predicted and no experimental realizations were yet performed. Fortunately, two main features were of substantial help in experimentally verifying this phenomenon: 1)  a thermal source can be easily simulated in a laboratory by using laser light impinging on a fast rotating ground glass \cite{Arecchi196627}; 2) given a coherence time for a thermal source which can range from the order of $ns$ to $\mu s$, the calibration of the interferometric paths and of the detection times  can be easily achieved. Furthermore, the same effect can be realised also in the spatial domain, by substituting the two Mach-Zehnder interferometers with two double slits and performing spatially correlated measurements at the output, as proposed in Ref. \cite{Cassano}.  Indeed, it did not take too long until experimental  demonstrations of this effect were first reported both in the temporal and spatial domain in Refs. \cite{Kim,DAngelo}, respectively.  Furthermore, as predicted in Ref. \cite{Tam-Sei}, interference fringes as a function of the phase difference $\varphi_{C}-\varphi_{T}$ have been observed with almost constant visibility at the increasing of the path length difference $L-S$ even if of the order of $680 m$ beyond the coherence length of the source \cite{Kim}.  This  phenonenon therefore paves the way to applications in high precision metrology and sensing, including the characterization of remote objects as demonstrated first theoretically in Ref. \cite{Cassano} and than experimentally in Ref. \cite{DAngelo}.  

In addition, the genuineness of the \textit{correlations} between the long and the short paths in the observed second order interference is testified by the possibility to exploit them to simulate quantum logic gates in absence of entanglement \cite{Tam-Sei}. Indeed, the simulation of a CNOT gate by using this interference effect, first theoretically proposed in the temporal domain in Ref. \cite{Tam-Sei} and extended to the spatial domain in Ref. \cite{Cassano}, was experimentally realised in Ref. \cite{Peng}. This can lead to potential applications in the development of novel optical algorithms with thermal light \cite{Wolk2011,tamma_analogue_2015,tamma_analogue_2015-1,PRARapidFact,TamAll}.

In this paper, we provide a deeper understanding of the physics behind this interference phenomenon in terms of  interference of ``two-photon detection amplitudes'' \cite{Glauber1963,GlauberLecture}. We will demonstrate how, differently from standard HBT experiments, interference can actually occur in a phase-dependent manner between two-photon detection amplitudes corresponding to the propagation of a pair of detected photons $(m,n)$ from the source to the detectors $d=,C,T$  through two possible pairs  $(L_C,S_T)$ and $(L_T,S_C)$  of  two different types of paths $L$ and $S$  (Fig. 2(c)). Further, standard HBT interfering amplitudes also arise from the propagation through pairs of  the same type of path (either $(L_C,L_T)$ interfering with $(L_T,L_C)$ or $(S_C,S_T)$ interfering with $(S_T,S_C)$ (Fig. 2(b)).

It is than the combination of all these  interference contributions from all the possible pairs of photons emitted by the source that leads  to the observed ``collective''  interference between the pair $(L_C,L_T)$ of long paths and the pair $(S_C,S_T)$  of short paths. Evidently, this does not mean  that each pair $(m,n)$ of two detected photons have either taken the $(L_C,L_T)$ or the $(S_C,S_T)$ paths in an indistinguishable manner. Indeed, the correlated paths $(S_C,S_T)$ and $(L_C,L_T)$ \textit{cannot} be interpreted as \textit{two-photon} paths. On the other hand, their interference is a result of the non trivial \textit{collective} contribution of all the possible pairs $(m,n)$ of detected  photons associated with different interfering two-photon detection amplitudes within the statistics of the thermal source (Figs. 2(b) and 2(c)).

In Section  \ref{sec1}, we will first analyze this  interference phenomenon  by using the standard description of thermal light in terms of the Glauber-Sudarshan probability distribution (Eq. (\ref{Pfunction})). In Section \ref{sec2},  we will give a detailed analysis of the fundamental two-photon interference physics underlying this effect. We will finally conclude the paper with conclusions and perspectives in Section \ref{sec3}.

\section{Thermal light interference ``beyond coherence''}\label{sec1}
We consider the interferometer in Fig. 1(a), with  the input thermal state described by  \cite{Glauber2007,Mandel1995}
\begin{align}
 \hat{\rho}_{ther}
  & =
  \int \left[\prod_{\omega} \difd^2 \alpha_{\omega}\right] P_{\hat{\rho}_{ther}}(\{\alpha_{\omega}\}) \bigotimes_{\omega} \ket{\alpha_{\omega}} \! \bra{\alpha_{\omega}} 
\label{eq:Density_op_Thermal_B_Two_Mode},
\end{align}
with the Glauber-Sudarshan probability distribution \cite{Glauber1963, Sudarshan1963}
\begin{equation}
 P_{\hat{\rho}_{ther}}(\{\alpha_{\omega}\})=\prod_{\omega}\frac{1}{\pi\,\overline{n}_{\omega}}\exp\left(-\frac{\abssq{\alpha_{\omega}}}{\overline{n}_{\omega} }\right),
 \label{Pfunction}
\end{equation}
and the average photon number  \cite{Glauber2007}
\begin{align}
  \overline{n}_{\omega} &= \overline{r} \frac{1}{\sqrt{2\pi}\Delta \omega} \, \exp\left\lbrace -\frac{\left(\omega - \omega_{0} \right)^2}{2 \Delta\omega^{2}} \right\rbrace
  \nonumber
\label{eq:C02_014_Text},
\end{align}
at frequency $\omega$, with the mean photon rate $\overline{r}$, average frequency $\omega_{0}$ and spectral width $\Delta \omega$.

At the interferometer output the intensity fluctuations
\begin{equation}
\Delta I_{d} (t_d) = I_{d} (t_d) - \avg{I_{d} (t_d)}
\label{eq:pnf_def}
\end{equation}
at the detection time $t_{d}$ around the mean value $\avg{I_{d} (t_d)}$, at each detector $d=C,T$, can be measured by using single-photon detectors with integration time $\delta t \ll 1/\Delta \omega$ \cite{ShihScully2014, Shih}.
One can than evaluate the  correlation in the intensity fluctuations at the two output ports
\cite{ShihScully2014, Shih, Glauber2007}
\begin{align}
\avg{\Delta I_{C}\Delta I_{T}} &= \avg{I_{C}\,I_{T}} - \avg{I_{C}}\avg{I_{T}}\notag \\ 
\label{eq:corrfluc}
\end{align}
by subtracting from the correlation in the intensities  \cite{Glauber2007}
\begin{equation}
\avg{I_{C}(t_C) \, I_{T}(t_T)} \propto G^{(2)}(t_{C},t_{T}) = \tr \left[\hat{\rho}_{ther} {\hat{E}}_{C}^{(-)}\!(t_{C})  {\hat{E}}_{T}^{(-)}\!(t_{T}) {\hat{E}}_{C}^{(+)}\!(t_{C}){\hat{E}}_{T}^{(+)}\!(t_{T})  \right], 
\label{PhotonNumberCorrelations}
\end{equation}
in terms of the field operators $ {\hat{E}}_{d}^{(+)}\!(t_{d})$ and their Hermitian conjugate $ {\hat{E}}_{d}^{(-)}\!(t_{d})$ at the detectors $d=C,T$, the background term \cite{Tam-Sei}
\begin{equation}
\avg{I_{C} (t_C)} \avg{I_{T}(t_T)}\propto G^{(1)}(t_{C},t_{C}) G^{(1)}(t_{T},t_{T}) = 2a^2\overline{r}^2, 
\label{background}
\end{equation}
associated with the product of the intensities
\begin{equation}
G^{(1)}(t_{d},t_{d})\defeq\tr \left[\hat{\rho}_{ther} {\hat{E}}_{d}^{(-)}\!(t_{d})  {\hat{E}}_{d}^{(+)}\!(t_{d})  \right],
\label{eq:first-order-correlation}
\end{equation}
where $a$ is a constant.

At approximately equally detection times with respect to the coherence time $1/\Delta\omega$ of the source,  
 \begin{align}
 \abs{t_{C}-t_{T} }&\ll 1/\Delta\omega \label{eq:Cnot conditions 01c}
\end{align}
and for path delays 
 \begin{align}
\abs{L_{C}-L_{T}}/c \,\,, \abs{S_{C}-S_{T}}/c  \ll 1/\Delta\omega,   \quad\quad\label{eq:Cnot conditions 02a} 
\quad\quad
\abs{L_{T}-S_{T}}/c \,\,, \abs{L_{C}-S_{C}}/c \gg 1/\Delta\omega,
\end{align} 
the expression in Eq. (\ref{eq:corrfluc}) becomes \cite{Tam-Sei}
\begin{align}
\avg{\Delta I_{C} \Delta I_{T}}
&\propto \abssq{G^{(L_{C},L_{T})}(t_{C},t_{T})+G^{(S_{C},S_{T})}(t_{C},t_{T})}
\nonumber \\ 
&= 2a^2\overline{r}^2 \left(1+ \cos (\varphi_{C}-\varphi_{T})   \right).
\label{eq:pnf coh gen}
\end{align}
Here, the interference between the two-path contributions \cite{Tam-Sei}
\begin{equation}
G^{(l_{C},l_{T})}(t_{C},t_{T}) = \ii \, a\,\overline{r}\, \e^{\ii \omega_{0}[t_{C}-t_{T}]}\e^{- \ii \omega_{0}[l_C- l_T]/c}\e^{[l_C- l_T]^2 \Delta \omega^2 /2c^2}\e^{-[t_C- t_T]^2 \Delta \omega^2 /2},
\label{eq:Gll}
\end{equation} 
with $(l_{C},l_{T})= (L_{C},L_{T}),(S_{C},S_{T})$, leads to a sinusoidal dependence on the \textit{difference} $\varphi_{C}-\varphi_{T}$ between the relative phases $\varphi_{d}=\omega_{0}(L_{d}-S_{d})/c $, with $d=C,T$, in the two Mach-Zehnder interferometers. These sinusoidal oscillations can be measured independently of how much the difference in length between the L-type paths and the S-type path is beyond the coherence length of the source, since each interfering contribution in Eq. (\ref{eq:pnf coh gen}) depends only on the relative distance between paths of the same type. Indeed, the other interfering terms $G^{(S_{C},L_{T})}(t_{C},t_{T})$ and $G^{(L_{C},S_{T})}(t_{C},t_{T})$ from path of different types give a negligible contributions to Eq. (\ref{eq:pnf coh gen}) in the conditions in Eq. (\ref{eq:Cnot conditions 02a}). 

We point out that the phase difference $\varphi_C - \varphi_T = \omega_{0}[(L_C - L_T) - (S_C - S_T)]/c$ can be tuned by varying the difference in length  between paths of the same type $(0 \leq \omega_{0}(l_C - l_T)/c \leq 1)$, within the coherence length of the source as in Eq. (\ref{eq:Cnot conditions 02a}). This was obtained, for example, experimentally in Ref. \cite{Kim} by using a thermal source of coherence length of about $120 \,  m$ and central  wavelength of $780 \, nm$ by varying the differences in the path lengths with piezoelectric actuators at the rate of $63 \, nm/s$. We also emphasize that interference fringes can be observed also by measuring directly the correlation in the photon numbers in Eq. (\ref{PhotonNumberCorrelations}) but with a visibility of $1/3$ due to the additional background term in Eq. (\ref{background}).

How can we interpret these second-order interference fringes in terms of two-photon interference?

\section{Two-photon interference physics}\label{sec2}

We describe here the observed interference effect in terms of interference of two-photon detection amplitudes \cite{Glauber1963,GlauberLecture}.
Toward this goal, it useful to mimic the thermal field emitted from the point-like source by a very large number of subfields m with frequency $\omega_m$, which state can be written in the coherent representation as  \cite{ShihBook}
\begin{equation}
|\Psi \rangle \doteq \prod_{m} |\alpha_m({\omega_m})\rangle,
\label{alphastate}
\end{equation}
where $|\alpha_m({\omega_m})\rangle$ is an eigenstate of the annihilation
operator  $\hat{a}_m({\omega_m})$ with eigenvalues $\alpha_{m}({\omega_m})$ which contain a real-positive amplitude $a_m({\omega_m})$ and a random phase $\varphi_m({\omega_m})$ arising from the thermal nature of the subfields emitted by the source.  

We can then evaluate the correlation in the intensity fluctuations 
\begin{align}\label{Photon_number_correlation}
\big{\langle} \Delta I_C(t_C)  \Delta I_T(t_T)\big{\rangle}&\propto\big{\langle}\langle\Psi|\hat{E}^{(-)}(t_C)\hat{E}^{(-)}(t_T)\hat{E}^{(+)}(t_C)\hat{E}^{(+)}(t_T)|\Psi\rangle\big{\rangle}_{Es}  \nonumber\\
&- \big{\langle}\langle\Psi|\hat{E}^{(-)}(t_C)\hat{E}^{(+)}(t_C)|\Psi\rangle\big{\rangle}_{Es}
\big{\langle}\langle\Psi|\hat{E}^{(-)}(t_T)\hat{E}^{(+)}(t_T)|\Psi\rangle\big{\rangle}_{Es},
\end{align}
where $\langle\cdots\rangle_{Es}$ denotes the ensemble average over all the possible values of $\alpha_{m}({\omega_m})$. Here, the field operator $\hat{E}^{(+)}(t_d)$ at the detectors $d=C, T$ can be expressed as the sum
\begin{align}
\hat{E}^{(+)}(t_d)=\sum_m \sum_{\,\,\,\,\, \xrightarrow{}   =  \xdasharrow{\text{}}  ,  \xdotarrow{\text{}} }{\hat{E}^{(+)}_{m \xrightarrow{}  d }(t_d)},
\label{fields}
\end{align}
 where the subfield operators
\begin{equation}
{\hat{E}}_{m \xrightarrow{}  d }^{(+)}(t_{d})\propto\!\!\int\!\! \difd \omega\; \e^{-i\omega [t_{d} -  l_{m \xrightarrow{}  d }/c ] }\;
\hat{a}_m(\omega)
\end{equation}
take into account the  propagation of each detected photon m from the point-like source where each corresponding subfields m is generated  to the point-like detector $d=C,T$ through either the long path $l_{m \xdasharrow{\text{}}  d}= L_d$   or the short path $l_{m \xdotarrow{\text{}}  d} = S_d$.

We can now describe the correlation in the intensity fluctuations in Eq. (\ref{Photon_number_correlation}) in terms of the interference of all the possible ``two photon-detection amplitudes''
\begin{equation}
\begin{bmatrix}
         m  \xrightarrow{} C\\
         n  \xrightarrow{} T 
        \end{bmatrix} \doteq  \langle\alpha_m(\omega_m)|{\hat{E}}_{m \xrightarrow{}  C }^{(+)}(t_{C})|\alpha_m(\omega_m)\rangle \langle\alpha_n(\omega_n)|{\hat{E}}_{n \xrightarrow{}  T }^{(+)}(t_{T})|\alpha_n(\omega_n)\rangle,
        \label{amplitudes}
\end{equation}
with $\xrightarrow{}   =  \xdasharrow{\text{}}  ,  \xdotarrow{\text{}}$, associated with all the possible contributions of two different subfields m and n to a two-photon detection, in analogy to a HBT experiment. However, since the time delay between the long (dashed red) and the short (dotted blue) paths (Fig. 2(a)) is beyond the  coherence time of the source, interference can occur only if each given subfield  m,n  undertakes the same type of path in the two-interfering two-photon amplitudes. There are therefore only four possible ways for these amplitudes to interfere with each other as in the expression 
\begin{eqnarray}
\big{\langle} \Delta I_C(t_C)  \Delta I_T(t_T)\big{\rangle} &\propto
\big{\langle} \sum_{m\neq n} 
\begin{bmatrix}
         m   \xdasharrow{\text{}}  C\\
         n   \xdasharrow{\text{}}  T 
        \end{bmatrix}
       \begin{bmatrix}
         m   \xdasharrow{\text{}}  T\\
         n   \xdasharrow{\text{}}  C 
        \end{bmatrix}^*
        +
        \begin{bmatrix}
        m   \xdotarrow{\text{}}  C\\
         n   \xdotarrow{\text{}}  T 
        \end{bmatrix}
       \begin{bmatrix}
         m   \xdotarrow{\text{}}  T\\
         n   \xdotarrow{\text{}}  C 
        \end{bmatrix}^*
          \big{\rangle}_{Es}
\nonumber   \\     
                &+ \big{\langle} \sum_{m\neq n} 
        \begin{bmatrix}
        m   \xdasharrow{\text{}}  C\\
         n   \xdotarrow{\text{}}  T 
        \end{bmatrix}
       \begin{bmatrix}
         m   \xdasharrow{\text{}}  T\\
         n   \xdotarrow{\text{}}  C 
        \end{bmatrix}^* 
         + 
        \begin{bmatrix}
        m   \xdotarrow{\text{}}  C\\
         n   \xdasharrow{\text{}}  T 
        \end{bmatrix}
       \begin{bmatrix}
         m   \xdotarrow{\text{}}  T\\
         n   \xdasharrow{\text{}}  C 
        \end{bmatrix}^*    
        \big{\rangle}_{Es}\nonumber\\
        \label{corr}
\end{eqnarray}
of the correlation in the intensity fluctuations, which follows from Eq. (\ref{Photon_number_correlation}) by using Eqs.  (\ref{alphastate}, \ref{fields}, \ref{amplitudes}).
Here, the large number of contributing subfields m,n spans in a continuous way all the possible eigenvalues $\alpha_{\omega}=\alpha_m(\omega_m),\alpha_n(\omega_n)$ in Eq. (\ref{amplitudes}), which contribute to the ensemble average in Eq. (\ref{corr}) according to the probability distribution associated with the P function in Eq. (\ref{Pfunction}). One can therefore write the first two contributions in Eq. (\ref{corr}) in terms of the phase-independent terms $\abssq{G^{(L_{C},L_{T})}(t_{C},t_{T})}$ and  $\abssq{G^{(S_{C},S_{T})}(t_{C},t_{T})}$ (defined by Eqs. (\ref{eq:Gll})) as
\begin{eqnarray}
\big{\langle} \sum_{m\neq n} 
\begin{bmatrix}
         m   \xdasharrow{\text{}}  C\\
         n   \xdasharrow{\text{}}  T 
        \end{bmatrix}
       \begin{bmatrix}
         m   \xdasharrow{\text{}}  T\\
         n   \xdasharrow{\text{}}  C 
        \end{bmatrix}^*
        +
        \begin{bmatrix}
        m   \xdotarrow{\text{}}  C\\
         n   \xdotarrow{\text{}}  T 
        \end{bmatrix}
       \begin{bmatrix}
         m   \xdotarrow{\text{}}  T\\
         n   \xdotarrow{\text{}}  C 
        \end{bmatrix}^*
          \big{\rangle}_{Es}
          \nonumber\\    
          \simeq \abssq{G^{(L_{C},L_{T})}(t_{C},t_{T})}+\abssq{G^{(S_{C},S_{T})}(t_{C},t_{T})}
          = 2a^2\overline{r}^2 
          \label{Contribution1}.
\end{eqnarray}
These interference contributions arise from the interference of two-photon amplitudes associated with two paths of the same type (long or short) within the coherence length of the source. Evidently,  interference terms of this type arise in an analogous way also in standard HBT interferometers (Fig. 2(b)) where $(L_C,L_T)$ or $(S_C,S_T)$ define the path lengths from the output ports of the beam splitter to the two detectors. Since, in this case, the paths associated with the two corresponding interfering amplitudes completely overlap spatially, neither of these two terms can be sensitive to any phase delay, as expected for standard HBT interference.

On the other hand, by again using Eqs. (\ref{Pfunction},\ref{eq:Gll},\ref{amplitudes}), the last two interference contributions in Eq. (\ref{corr}) can be expressed as  (c.c. stands for complex conjugate) 
\begin{eqnarray}
\hspace*{-2.5cm} 
\big{\langle} \sum_{m\neq n} 
        \begin{bmatrix}
        m   \xdasharrow{\text{}}  C\\
         n   \xdotarrow{\text{}}  T 
        \end{bmatrix}
       \begin{bmatrix}
         m   \xdasharrow{\text{}}  T\\
         n   \xdotarrow{\text{}}  C 
        \end{bmatrix}^* 
         + 
        \begin{bmatrix}
        m   \xdotarrow{\text{}}  C\\
         n   \xdasharrow{\text{}}  T 
        \end{bmatrix}
       \begin{bmatrix}
         m   \xdotarrow{\text{}}  T\\
         n   \xdasharrow{\text{}}  C 
        \end{bmatrix}^*    
        \big{\rangle}_{Es} =
\big{\langle} \sum_{m\neq n} 
        \begin{bmatrix}
        m   \xdasharrow{\text{}}  C\\
         n   \xdotarrow{\text{}}  T 
        \end{bmatrix}
       \begin{bmatrix}
         m   \xdasharrow{\text{}}  T\\
         n   \xdotarrow{\text{}}  C 
        \end{bmatrix}^* 
        + c.c.       
        \big{\rangle}_{Es} 
        \nonumber\\
       \simeq[G^{(L_{C},L_{T})}]^* G^{(S_{C},S_{T})}(t_{C},t_{T}) +c.c.
       = 2a^2\overline{r}^2 \cos (\varphi_{C}-\varphi_{T}) .
       \label{Contribution2}
\end{eqnarray}
Here, the two detected photons either have undertaken the $(L_C,S_T)$ path or the $(L_T,S_C)$ path, corresponding to pairs of two paths of different types (Fig. 2(c)),  due to the absence of ``which path information''  in the propagation after the beam splitter.  Therefore, these terms are sensitive to the phase difference $\varphi_{C}-\varphi_{T}$  between the two unbalanced interferometers leading to the observation of sinusoidal fringes, which are unique to this interference phenomenon with respect to standard HBT interference. This is also very different from Franson-type interference where the two photons either take the $(L_C,L_T)$ or the $(S_C,S_T)$ path, leading to a dependence on the sum $\varphi_{C}+\varphi_{T}$ of the two interferometer relative phases.

Interestingly, one can notice that it does not matter how far the long and short paths are apart from each other beyond the coherence length of the source since each subfields m,n always evolve through two possible paths of the same type (either $L=L_C,L_T$ or $S=S_C,S_T$) which length difference is within the coherence length of the source. Therefore the corresponding interfering two photon-amplitudes are always indistinguishable.     Of course, the larger is the path unbalance in each Mach Zehnder interferometer the larger has to be the emission time delay between the photons m and n associated with the corresponding subfields triggering the two-fold detection at approximately equal detection times. However, given the continuous nature of the thermal field there is not, in principle, any constraint to such a time delay.  This is evidently very different from the situation in the Franson interferometer where the time delay associated with the interferometer unbalance is required to be within the the coherence time of the pump laser used to generate the entangled pair of photons. 

Remarkably, the ``incoherent'' sum in Eq. (\ref{corr}) of the two-photon interference contributions in Eqs. (\ref{Contribution1}) and (\ref{Contribution2}) leads to the interference of the two ``effective'' two-path amplitudes associated with the two pairs $(L_{C},L_{T})$ and $(S_C,S_T)$ of correlated paths as in Eq.  (\ref{Photon_number_correlation}). However, these are \textit{not} interfering two-photon detection amplitudes, since no  pairs of photon can have undertaken these two pairs of paths, with length difference $L-S$ beyond the coherence length of the source, in an indistinguishable manner.  On the other hand, such two-path interference  is  a ``collective'' result of all the possible two-photon interference contributions in Eq. (\ref{corr}) as depicted in Figs. $2(b)$ and $2(c)$.

\section{Conclusions and perspectives}\label{sec3}
We have described here how thermal light second-order interference beyond the coherence time arises, for any pair (m,n) of detected photons, from the interference of two-photon detection amplitudes without contradicting our understanding of two-photon coherence.

In particular, for each pair of interfering two-photon detection amplitudes,  each detected photon can undertake only paths of the same type (either L or S). Indeed, only in this case the corresponding path differences $L_C-L_T$ or $S_C,-S_T$ are within the coherence length of the source (Figs. 2(b) and 2(c)).

If the m photon undertakes always the same type of path as the n photon before triggering a two-fold detection (Fig. 2 (b)) no sinusoidal fringes could be observed.  These ``HBT interference contributions'' are usual in standard HBT experiments (Fig. 1(b)). However, the novelty here is that the (m,n) pair of indistinguishable photons can also undertake indistinguishable pairs of different types of paths, $(L_C,S_T)$ interfering with $(L_T,S_C)$. Remarkably, these ``non-HBT interference contributions'' are sensitive to difference   $\varphi_C-\varphi_T$ between the phases in the two unbalanced interferometer. Evidently, in this case, the emission time delay between the two photons is of the order of $(L-S)/c$ which is beyond the coherence time of the source. However, since the thermal field is a continuous field, there is no limit, in principle,  to such a time delay and therefore to the path unbalance in each Mach-Zehnder interferometer.

Interestingly, even if the described HBT and non-HBT contributions add incoherently,  an effective interference between the two pairs $(L_{C},L_{T})$ and $(S_C,S_T)$ of correlated paths arises from the collective contribution of all the possible pairs of photons emitted by the thermal source. This two-path interference is evidently beyond the thermal light coherence length. One can think that even if two-photon interference occurs within the coherence length of the source, a ``collective'' second-order interference can occur beyond such coherence length. In other words, the ``incoherent'' sum of all possible two-photon interference contributions can generate ``coherence'' in a collective manner according with the statistics of the thermal source.

We emphasize that  the same physics described here naturally applies also to experiments in the spatial domain where the two Mach Zehnder interferometers in Fig. 2(a) are substituted by two double slits \cite{Cassano,DAngelo}.

\ack

Research
was partially sponsored by the Army Research Laboratory and was accomplished
under Cooperative Agreement Number W911NF-17-2-0179. The views and conclusions
contained in this document are those of the authors and should not be interpreted
as representing the official policies, either expressed or implied, of the Army
Research Laboratory or the U.S. Government. The U.S. Government is authorized
to reproduce and distribute reprints for Government purposes notwithstanding any
copyright notation herein.

V.T is also thankful to Kurt Jacobs for useful discussions.

This work is dedicated to Professor Wolfgang P. Schleich on the occasion of his 60th birthday. Indeed, the interference effect at the heart of this work was first predicted at the Institute of Quantum Physics in Ulm together with Johannes Seiler, who completed his bachelor thesis under my supervision. It was thanks to Wolfgang's encouragement and advice that the description of this phenomenon was finally published in New Journal of Physics as a Fast Track Communication \cite{Tam-Sei}. I am extremely thankful to him for believing in me and supporting me in all my  research even before I joined him in Ulm.  Wolfgang has been and will always be a source of inspiration for me, not only because he is an outstanding scientist but also because he is dedicated to supporting young scientists and valuing their research. Thank you Wolfgang for your friendship and for being such a wonderful mentor for me! Happy birthday!

\section*{ORCID iD}
V. Tamma  https://orcid.org/0000-0002-1963-3057

\section*{References}


\begin{thebibliography}{99}

\bibitem{Tam-Sei}
V. Tamma and J. Seiler,  ``Multipath correlation interference and controlled-not gate simulation with a thermal source,'' New
J.  Phys. {\bf 18}, 032002 (2016).

\bibitem{HBT1}
 R. Hanbury Brown and R.Q. Twiss, ``Correlation between photons in two coherent beams of light,''   Nature {\bf 177}, 27 - 29 (1956).
  
\bibitem{HBT2}
R. Hanbury  Brown and R.Q. Twiss, ``A test of a new type of stellar interferometer on sirius,'' Nature {\bf 178}, 1046 -1048 (1956).

\bibitem{Schleich}
W. P. Schleich, `` Quantum Optics in Phase Space'' (Berlin:
Wiley-VCH, 2001).

\bibitem{Glauber2007}
{R.J.} Glauber  {\em {Quantum Theory of Optical Coherence: Selected Papers and  Lectures}\/} (John Wiley and Sons, 2007).

\bibitem{Scully1997}
{M.O.} Scully, and {M.S.} Zubairy,   {\em {Quantum Optics}\/} (Cambridge University Press, 1997).

\bibitem{ShihBook}
{Y.H.} Shih {\em {An Introduction to Quantum Optics}\/} (CRC Press Taylor and Francis group, 2011).

\bibitem{Mandel1995}
{L.} Mandel and {E.} Wolf {\em Optical Coherence and Quantum Optics\/} (Cambridge University Press, 1995).


\bibitem{Glauber1963}
R.J. Glauber,  ``Photon correlations,'' {Phys. Rev. Lett.} {\bf 10}, 84 (1963).
  
\bibitem{GlauberLecture} 
R.J. Glauber,  ``One hundred 
{years} of light quanta,'' Nobel Lecture {8 Dec 2005, (The Nobel Foundation 2005)}.

\bibitem{ShihAlley}
C.O.Alley  and Y.H.Shih  {\em Proceedings of the Second International
  Symposium on Foundations of Quantum Mechanics in the Light of New
  Technology\/} ed of~Japan P~S (Tokyo, 1986), pp 47 - 52;
\nonumber\\
Y. H. Shih and C. O. Alley, ``New type of Einstein-Podolsky-Rosen-Bohm experiment using pairs of light quanta
produced by optical parametric down conversion,'' Phys. Rev. Lett. {\bf 61}, 2921 (1988).

\bibitem{HOM}
C. K. Hong, Z. Y. Ou, and L. Mandel, ``
{Measurement} of subpicosecond time intervals betweens two photons by
interference,'' Phys. Rev. Lett. {\bf 59}, 2044 (1987).

\bibitem{Franson1989}
J~D Franson, ``Bell inequality for position and time,'' {\em Phys. Rev. Lett.\/} {\bf 62}(19) 2205--2208 (1989).

\bibitem{Legero}
T. Legero, T. Wilk, M. Hennrich, G. Rempe, and A. Kuhn, ``Quantum Beat of Two Single Photons,'' Phys. Rev. Lett. {\bf 93}, 070503 (2004).

\bibitem{HOMspatial}
 H. {Kim},  O. {Kwon}, W. {Kim}, {and} T. {Kim}, ``Spatial two-photon interference in a Hong-Ou-Mandel interferometer,'' {Phys.  Rev. A} {\bf 73}, 023820 (2006).
 
\bibitem{LiuSpatial}
J. {Liu}, Y. {Zhou}, W. {Wang},  R.F. {Liu}, K. {He}, F.L. {Li}{,}  and Z. {Xu}, ``Spatial second-order interference of
pseudothermal light in a Hong-Ou-Mandel interferometer,''  {Optics Express} {\bf  21}(16), 19209 - 19218 (2013).

\bibitem{tammalaibacher2014a}
V. Tamma and S. Laibacher, ``Multiboson correlation interferometry with multimode thermal sources,'' Phys. Rev. A {\bf 90},
063836 (2014).

\bibitem{tammalaibacher2014}
V. Tamma and S. Laibacher, ``Multiboson correlation interferometry with arbitrary single-photon pure states''  Phys. Rev. Lett. {\bf 114}, 243601 (2015).

\bibitem{Kim}
Y.~S. {Ihn}, Y. {Kim}, V. {Tamma}, Y.-H.{Kim}, 
``Second-Order Temporal Interference with Thermal Light: Interference beyond the Coherence Time,'' Phys. Rev. Lett. {\bf 119}, 263603 (2017).

\bibitem{Genov}
{M. Genovese ``Real applications of quantum imaging,'' Journal of Optics {\bf 18}(7), 073002 (2016).}


\bibitem{Lemos}
{G.B. Lemos, V. Borish, G.D. Cole, S. Ramelow, R. Lapkiewicz, and A. Zeilinger, ``Quantum imaging with undetected photons,'' Nature {\bf 512}, 409–412 (2014).}

\bibitem{Jane}
{J. Sprigg, T. Peng, and Y.H. Shih , ``Super-resolution imaging using the spatial-frequency filtered intensity fluctuation correlation,'' Scientific Reports {\bf 6}, 38077 (2016).}

\bibitem{PittShi}
T.B. Pittman, Y.H. Shih,  D.V. Strekalov{,} and  A.V. Sergienko,  ``Optical imaging by means of two-photon quantum entanglement,'' {Phys. Rev. A} {\bf 52}, R3429 (1995).

\bibitem{ValScaD'AnShi}
A. Valencia, G. Scarcelli, M. D'Angelo, and Y.H. Shih ``Two-photon imaging with thermal light,'' {Phys. Rev. Lett.} {\bf 94}, 063601 (2005).

\bibitem{D'AnShi} 
M. {D'Angelo} and Y.H. {Shih},``Quantum Imaging,'' Laser Phys. Lett. {\bf 2}{(12)}, 567 {- 596} (2005).
  
\bibitem{Scarcelli} 
G. {Scarcelli}, V. {Berardi}{,}  and Y.H. {Shih}, ``Can two-photon correlation of chaotic light be considered as correlation of intensity fluctuations?,'' {Phys. Rev. Lett.} {\bf 96}, 063602 (2006).

\bibitem{gatti2}
A. Gatti, E. Brambilla, M. Bache, and L. A. Lugiato, ``Correlated imaging, quantum and classical,'' Phys. Rev. A {\bf 70}, 013802 (2004).

\bibitem{ferri}
F. Ferri, D. Magatti, A. Gatti, M. Bache, E. Brambilla, and L. A. Lugiato, ``High-resolution ghost image and ghost diffraction experiments with thermal light,''  Phys. Rev. Lett. {\bf 94}, 183602 (2005).

\bibitem{Chen}
H. Chen, T. Peng{,}  and Y.H. Shih, ``100\% correlation of chaotic thermal light,'' Phys. Rev. A {\bf 88}, 023808 (2013).

\bibitem{Luo}
K.-H. Luo, B.-Q. Huang, W.-M. Zheng and L.-A.Wu ``Nonlocal Imaging by Conditional Averaging of Random Reference Measurements,'' Chin. Phys. Lett. {\bf 29}{(7)}, 074216 (2012).

\bibitem{Kok}
{P. Kok, ``Photonic quantum information processing,'' Contemporary Physics {\bf 57}(4), 526 - 544 (2016).}

\bibitem{nielsen}
M. Nielsen and I. Chuang {\em Quantum Computation and Quantum Information\/} (Cambridge Series on Information and the Natural Sciences, Cambridge  University Press, 2000).

\bibitem{BosonSampling} S. Aaronson and A. Arkhipov, in Proceedings of the forty-third annual ACM symposium on Theory of
computing (ACM, New York), pp. 333–342 (2011).

\bibitem{laibachertamma2015} 
S. Laibacher and V. Tamma,  ``From the physics to the computational complexity of multiboson correlation interference,'' Phys. Rev. Lett. {\bf 115}, 243605 (2015).
 
\bibitem{tammalaibacher2015b}
V. Tamma and S. Laibacher,``Multi-boson correlation sampling,''  {Quantum Inf. Process.} {\bf 15}{(3)}, 1241 - 1262 (2016).

  
\bibitem{tammalaibacher2015}
V. Tamma and S. Laibacher, ``Boson sampling with non-identical single photons,'' {J. Mod. Opt.} {\bf 63}, 41 (2015).
  
\bibitem{GioLloMac1}
{V. Giovannetti, S. Lloyd, and L. Maccone,  ``Advances in quantum metrology,''  {Nature Photonics} {\bf 5}(4), 222-229 (2011).}

\bibitem{GioLloMac}
{V. Giovannetti, S. Lloyd, and L. Maccone,  ``Quantum-Enhanced Measurements: Beating the Standard Quantum Limit,''  {Science} {\bf 306}(5700), 1330-1336 (2004).}

\bibitem{Dowling} 
J. {Dowling}, ``Quantum optical metrology - the lowdown on high-N00N states,'' {Contemporary Physics} {\bf 49}{(2), 125 - 143} (2008).

\bibitem{Cassano}
M. Cassano, M. D'Angelo, A. Garuccio, T. Peng, Y.H. Shih and V. Tamma,  ``Spatial interference between pairs of disjoint optical paths with a single chaotic source,'' Opt. Express {\bf 6}, 6589--6603 (2017).

\bibitem{DAngelo}
M. D'Angelo, A. Mazzilli, F. V. Pepe, A. Garuccio,  and V. Tamma, ``Characterization of two distant double-slit by chaotic light second-order interference,'' Scientific Reports {\bf 7}, 2247 (2017).

\bibitem{Dowling2}
J.~P. Olson, K.~R. Motes, P.~M. Birchall, N.~M. Studer, M.~LaBorde, T.~Moulder,
  P.~P. Rohde and J.~P. Dowling, {\em Phys. Rev. A} {\bf 96}  (2017) p. 013810.

\bibitem{Zimmermann} 
O. { Zimmermann}  and V.  {Tamma}, `` Which role does multiphoton interference play in small phase estimation in quantum Fourier transform interferometers?,'' Int. J. Quantum Inf. {\bf 15}, 1740020 (2017).

\bibitem{NOON} 
M. {D'Angelo}, A.  {Garuccio}{,}  and V.  {Tamma}, ``Toward real maximally path-entangled  N -photon-state sources,'' {Phys.  Rev. A} {\bf 77}, 063826 (2008).

\bibitem{Arecchi196627}
Arecchi F, Gatti E and Sona A 1966 {\em Physics Letters\/} {\bf 20} 27 -- 29
  ISSN 0031-9163

\bibitem{Peng}
T. Peng, V. Tamma, and Y.H. Shih, ``Experimental controlled-not gate simulation with thermal light,'' Scientific Reports {\bf 6},
30152 (2016).

\bibitem{Wolk2011}
S. W\"olk, W. Merkel, W.P. Schleich, I.S. Averbukh{,} and B. Girard,  ``Factorization of numbers with Gauss sums: I. Mathematical background,'' New J. Phys.  {\bf 13}, 103007 (2011).

\bibitem{tamma_analogue_2015}
V. Tamma, ``Analogue algorithm for parallel factorization of an exponential number of large integers: I. theoretical description,'' Quantum Information Processing {\bf{15}}{(12), 5259 - 5280} (2015).

\bibitem{tamma_analogue_2015-1}
V. Tamma, ``Analogue algorithm for parallel factorization of an exponential number of large integers: II. optical implementation,'' Quantum Information Processing  { \bf{15}}{(12), 5243 - 5257} (2015).

\bibitem{PRARapidFact}
V. Tamma, H. Zhang, X. He, A. Garuccio, W.P. Schleich{,} and Y.H. Shih,  ``Factoring numbers with a single interferogram,'' Phys. Rev. A {\bf 83}, 020304 (2011).

\bibitem{TamAll}
{V. Tamma, C.O. Alley, W.P. Schleich {,} and Y.H. Shih, ``Prime Number Decomposition, the Hyperbolic Function and Multi-Path Michelson Interferometers,'' {Foundations of Physics} {\bf 42}(1), 111 - 121 (2012).}










%
%
%
%
%
%
%
%
%
%



\bibitem{Sudarshan1963}
{E.C.G.} Sudarshan, ``Equivalence of semiclassical and quantum mechanical description of statistical light beams,''  {Phys. Rev. Lett.} {\bf 10}, 277  (1963).







  
%
%
\bibitem{ShihScully2014}
Peng T, Chen H, Shih Y and Scully M~O 2014 {\em Phys. Rev. Lett.\/} {\bf
  112}(18) 180401
 

\bibitem{Shih}
Chen H, Peng T and Shih Y 2013 {\em Phys. Rev. A\/} {\bf 88}(2) 023808


%
\end{thebibliography}
\end{document}